\documentclass[prb,showpacs,floatfix,amsfonts]{revtex4}
\usepackage{graphicx,graphics,color,epsfig}% Include figure files
\usepackage{bm}
\usepackage{amsmath}
\usepackage{amssymb}

\newcommand{\bd}{{\bf d}}
\newcommand{\bB}{{\bf B}}

\newcommand{\bk}{{\bf k}}

\newcommand{\br}{{\bf r}}

\newcommand{\bA}{{\bf A}}
\newcommand{\ba}{{\bf a}}
\newcommand{\bJ}{{\bf J}}
\newcommand{\bE}{{\bf E}}

\newcommand{\beqa}{\begin{eqnarray}}
\newcommand{\eeqa}{\end{eqnarray}}

\renewcommand{\Re}{{\rm Re}}

\begin{document}
\preprint{}
\title{Excitonic condensate and quasiparticle transport in electron-hole
bilayer systems}
\author{Yogesh N. Joglekar$^{1}$, Alexander V. Balatsky$^{1}$, and
Michael P. Lilly$^{2}$}
\affiliation{$^1$ Theoretical Division, Los Alamos National Laboratory,
Los Alamos, New Mexico 87545\\ $^2$ Sandia National Laboratories,
Albuquerque, New Mexico 87185}
\date{\today}
\begin{abstract}
Bilayer electron-hole systems undergo excitonic condensation when the
distance $d$ between the layers is smaller than the typical distance between
particles within a layer. All excitons in this condensate have a fixed dipole
moment which points perpendicular to the layers, and therefore this
condensate of dipoles couples to external electromagnetic fields. We study
the transport properties of this dipolar condensate system based on a
phenomenological model which takes into account contributions from the
condensate and quasiparticles. We discuss, in particular, the drag and
counterflow transport, in-plane Josephson effect, and noise in the in-plane
currents in the condensate state.
\end{abstract}
\maketitle

%-----------------------------------------------------------------------------%

\section{Introduction}
\label{sec:intro} Bilayer quantum well systems, where (equilibrium) carriers in
one layer are electrons and the other layer are holes, have been
investigated extensively.~\cite{keldysh,bilayer,rice,shev,vig,palo} These
systems are one of the promising candidates for observing
Bose-Einstein condensation of excitons, the bound states of
electron-hole pairs.~\cite{sivan,kane,vij,pohlt} When the
distance $d$ between layers is small compared to the typical
distance $r_s$ between particles within each layer, excitons,
resulting from the attractive Coulomb interaction between
electrons and holes, form a dilute Bose gas and undergo
condensation. Bilayer electron-hole systems have the added
advantage that each exciton has a fixed dipole moment, which points in the
same direction for all excitons. Therefore, we call this excitonic condensate
a dipolar condensate \cite{ayp}.  This property, which is unique to
electron-hole bilayers, enables us to probe the
nominally neutral condensate via electric and magnetic fields
which are confined between the two layers.~\cite{ayp} In this sense, the
dipolar condensate resembles the A-phase of superfluid Helium-3, in which
all $^3$He Cooper pairs have fixed orbital moment that is pointing in the
same direction.~\cite{3he} Just as $^3$He is an orbital ferromagnet in the
A-phase, the electron-hole condensate is a ``ferromagnetic'' dipolar fluid,
in which all dipoles point in the same direction, are phase coherent, and
therefore give rise to for a nontrivial collective response to an
applied  magnetic field.

The formation and properties of excitonic condensates in {\it double quantum 
wells} is a subject of the ongoing debate. Although these systems are not truly
superfluid (the $U(1)$ symmetry associated with the phase of the exciton
order parameter is not an exact symmetry) various signatures of excitonic
condensation can be probed provided that the excitons are long-lived. At
present, there are three main candidates for realization of excitonic
condensation in double quantum wells: bilayers in which electron-hole plasma 
is created by optical pumping and then spatially separated by electric field 
to create indirect excitons, bilayer electron-electron quantum Hall systems 
near total filling factor one, and undoped bilayers in which carrier 
density in each layer can be independently controlled by an external gate 
associated with that layer. In the first system, the condensation
is detected {\it a posteriori} by photoluminescence
measurement~\cite{snoke,butov,lai} of electron-hole recombination and
transport measurements, which study the electromagnetic response of excitons, 
are not yet accessible.~\cite{voros} Therefore, an explicit demonstration of 
counterflow superfluidity in this system appears exceedingly difficult. 
The second system, quantum Hall bilayers, has been investigated by Eisenstein's
group~\cite{jpe} and Shayegan's group~\cite{shay}, and their remarkable
results provide promising signatures of excitonic condensation, albeit over
the vacuum of a fully-filled Landau level. Because of the non-trivial 
electromagnetic response of the ``vacuum'' underlying this condensate, the 
bilayer quantum Hall system is not a dipolar superfluid. In particular, it 
does not generate counterflow supercurrent in response to an in-plane
magnetic field.

Therefore, in this paper, we focus on the third candidate. Independently 
contacted electron-hole bilayers have been experimentally investigated in 
the past decade,~\cite{sivan,kane} albeit in the (high density) region 
where exciton condensation does not occur. It is simply a matter of not too 
distant time when we will have electron-hole bilayers with 
low density and high mobility, which support the realization of excitonic 
condensate. In this paper we address electronic signatures of such a 
condensate that allow for a diagnostic of the condensate state, including its 
superfluid properties. We emphasize that our predictions do not necessarily 
have counterparts in the quantum Hall bilayers,~\cite{jpe,shay} because the 
mapping of quantum Hall bilayers on to excitonic superfluid ignores the 
response of the fully-filled Landau level vacuum.

The plan of the paper is as follows. In the following section, we
recall basics of excitonic condensation in bilayer system. We
focus on transport experiments in which currents in the electron
and the hole layers are externally fixed, and the voltage drop
developed in each layer is measured. We calculate the resulting
electric field in each layer by using a two-fluid model. In
Sec.~\ref{sec:jj} we discuss the transport in bilayers in the
presence of a weak link. This system is closely related to a
Josephson junction. We present the results based on this analogy.
In Sec.~\ref{sec:noise} we present the spectrum of current noise
in these systems, in the absence of any external fields. We find
that the current noise can provide a measure of the superfluid
collective mode velocity. We conclude the paper with discussion
in Sec.~\ref{sec:disc}.

%-----------------------------------------------------------------------------%

\section{Transport}
\label{sec:transport}

Let us consider a bilayer system with holes in the top layer and
electrons in the bottom layer. We use a notation in which the hole
(electron) coordinates are given by $\br_h=\br+\bd/2$
($\br_e=\br-\bd/2$), $\br$ is a 2-dimensional position vector, and
$\bd=d\hat{z}$ is a vector normal to the
layers. The formation of excitonic condensate is signaled by a
nonzero value of the order parameter
$\Delta(\br)=\langle c^\dagger_{h}c^{\dagger}_{e}\rangle =
|\Delta|\exp[i\Phi(\br)]$, where $c^{\dagger}_{h}$ creates a hole
in the top layer and $c^{\dagger}_{e}$ creates an electron in the
bottom layer. At low energies, the
dipolar phase $\Phi(\br)$ is the only relevant degree of freedom.
In the presence of gauge potentials, the dipolar phase transforms
as $\nabla\Phi(\br)\rightarrow\nabla\Phi(\br)-e\ba(\br)$ where
$\ba(\br)=\bA_{e}(\br)-\bA_{h}(\br)$ is the difference between
vector potentials in the electron-layer and the hole-layer (We use
units such that $\hbar=1=c$). This leads to a dipolar supercurrent
in the condensate state, given by
$\bJ_d(\br)=2e\rho_d\left[\nabla\Phi(\br)-e\ba(\br)\right]$, where
$\rho_d$ is the dipolar phase-stiffness. For a smoothly varying
gauge potential, the antisymmetric combination $\ba\approx
d\partial_z\bA$ can be tuned by an in-plane magnetic
field; in particular, a uniform in-plane field $\bB_{||}=-B_{||}\hat{y}$ leads
to a uniform dipolar supercurrent
$\bJ_d=2e^2\rho_d\bd\times\bB_{||}$.~\cite{ayp}

\begin{figure}[thbp]
\begin{center}
\includegraphics[width=2in]{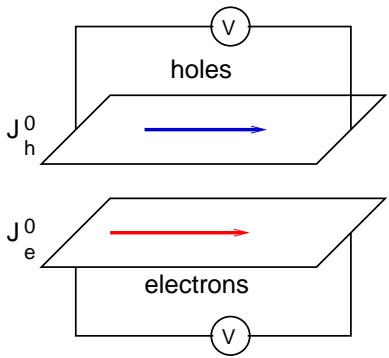}
\caption{(Color Online) Schematic electron-hole bilayer system, with fixed
current $J^0_h$ in the top (hole) layer and $J^0_e$ in the bottom
(electron) layer. We calculate the resulting electric fields (or
voltage drops) which develop in each layer by using a two-fluid
model.} \label{fig:ehcurrents}
\end{center}
\end{figure}

In this section, we consider transport experiments in which fixed
currents ${\bJ}^0_h$ and ${\bJ}^0_e$ flow through the hole-layer
and electron-layer, respectively. Our goal is to determine the
resulting electric fields (or voltage drops) in the individual
layers (Fig.~\ref{fig:ehcurrents}). In general the currents are
carried by the excitonic condensate and quasiparticles. We
characterize the quasiparticle contributions in the
linear-response regime as follows:
\begin{eqnarray}
\label{eq:jqph}
{\bJ}^{qp}_{h} & = &\sigma_{hh}{\bE}_{h}+\sigma_{he}{\bE}_{e},\\
\label{eq:jqpe} {\bJ}^{qp}_{e} & =
&\sigma_{ee}{\bE}_{e}+\sigma_{eh}{\bE}_{h}.
\end{eqnarray}
Here $\sigma_{ee},\sigma_{hh}$ are longitudinal conductivities and
$\sigma_{he},\sigma_{eh}$ are the drag conductivities. Adding the
condensate contribution gives the following equations for currents
in the two layers.
\begin{eqnarray}
\label{eq:j0h} {\bJ}^{0}_{h} & = &
+e\rho_d\left(\nabla\phi-e{\ba}\right)-\left(\sigma_{hh}
\partial_t{\bA}_{h}+\sigma_{he}\partial_t{\bA}_{e}\right),\\
\label{eq:j0e} {\bJ}^{0}_{e} & = &
-e\rho_d\left(\nabla\phi-e{\ba}\right)-\left(\sigma_{ee}
\partial_t{\bA}_{e}+\sigma_{eh}\partial_t{\bA}_{h}\right),
\end{eqnarray}
where we have used Maxwell's equation ${\bE}=-\partial_t{\bA}$. It
is convenient to rewrite the two equations, (\ref{eq:j0h}) and
(\ref{eq:j0e}), in terms of their sum and difference, leading to
the equations for antisymmetric gauge potential $\ba$ and the
symmetric gauge potential ${\cal A}=(\bA_h+\bA_e)$,
\begin{eqnarray}
{\bJ}^0_s & = & -(\sigma_s+\eta_s)\partial_t{\cal A}-(\sigma_d-\eta_d)
\partial_t{\ba},\\
{\bJ}^{0}_{d}& = &
+2e\rho_d(\nabla\Phi_d-e{\ba})-(\sigma_s-\eta_s)
\partial_t{\ba}-(\sigma_d+\eta_d)\partial_t{\cal A}.
\end{eqnarray}
Here, $\sigma_{s(d)}=(\sigma_{ee}\pm\sigma_{hh})$,
$\eta_{s(d)}=(\sigma_{he}\pm\sigma_{eh})$, and
$\bJ^{0}_s=(\bJ^{0}_h+\bJ^{0}_e)$ and
$\bJ^{0}_d=(\bJ^{0}_h-\bJ^{0}_e)$ are the external currents in the
symmetric and antisymmetric channel, respectively.

It is straightforward to solve these equations and obtain the time
dependence of vector potentials $\bA_h(t)$ and $\bA_e(t)$ within
each layer, provided that the supercurrent contribution is
stationary. The resulting electric fields in the two layers are
given by
\begin{eqnarray}
\label{eq:Eh} \bE_h(t)& =
&+\frac{\bJ^0_s}{2(\sigma_s+\eta_s)}+\frac{1}{2}(1-\sigma_{ds})
\alpha e^{-\alpha t}\tilde{\ba}(0),\\
\label{eq:Ee} \bE_e(t)& =
&+\frac{\bJ^0_s}{2(\sigma_s+\eta_s)}-\frac{1}{2}(1+\sigma_{ds})
\alpha e^{-\alpha t}\tilde{\ba}(0).
\end{eqnarray}
Here, $\sigma_{ds}=(\sigma_d-\eta_d)/(\sigma_s+\eta_s)$, the decay
rate
\begin{equation}
\alpha=\frac{2e^2\rho_d}{\left[(\sigma_s-\eta_s)-\sigma_{ds}(\sigma_d+\eta_d)
\right]},
\end{equation}
and we have defined
\begin{equation}
\tilde{\ba}=\frac{1}{2e^2\rho_d}\left[\bJ^0_s-2e\rho_d\nabla\Phi+\left(
\frac{\sigma_d+\eta_d}{\sigma_s+\eta_s}\right)J^0_d\right]
\end{equation}
which decays exponentially,
$\partial_t\tilde{\ba}(t)=-\alpha\tilde{\ba}(t)$. Decay rate
$\alpha$ is positive since typically $\sigma_s>\eta_s$ and
$\sigma_{ds}$ is small.  Eqs.(\ref{eq:Eh}) and (\ref{eq:Ee})
are the main results in this section. It follows that, at long
times $t\gg\alpha^{-1}$, electric fields in the two layers are
identical irrespective of individual currents flowing in each
layer.~\cite{vig} Their strength is determined by current in the
symmetric channel $\bJ^0_s$. This is consistent with the fact that
the dissipationless condensate contributes only to current in the
antisymmetric channel.

Now we turn our attention to two transport experiments.

i) First we focus on the drag setup, in which a fixed current
$J^0$ flows through the drive-layer and the voltage drop across
the second layer, namely the drag-layer, is measured,
$J^{0}_h=J^0$ and $J^0_e=0$. We find that the voltage in the drag
layer will be given by
\begin{equation}
\label{eq:vdrag} V_{drag}=\frac{LJ^0}{2(\sigma_s+\eta_s)}
\end{equation}
where $L$ is the linear sample size. When the system is deep in
the superfluid phase (when the temperature $T\rightarrow 0$ or
when $d/r_s\rightarrow 0$), the density of quasiparticles is
vanishingly small and therefore quasiparticle conductivity
vanishes. In that regime, therefore, $V_{drag}\rightarrow\infty$;
in other words, it is increasingly difficult to maintain a nonzero
current in the drive layer while keeping the drag layer open. We point 
out that the temperature dependence of the drag-voltage is determined 
by the temperature dependence of quasiparticle conductivities, and 
typically it will be different from the Coulomb drag which dominates 
when the system is not a dipolar superfluid.~\cite{rojo}

ii) The second experiment is the counterflow setup in which equal
and opposite currents flow through the two layers, $J^0_h=J^0$ and
$J^0_e=-J^0$. In this case, we find that the voltage drop across
either layers is zero. This is expected since, in the present
case, the current is carried entirely by the condensate and not by
the quasiparticles. We also find that the transient electric
fields in the two layers are given by
\begin{eqnarray}
\label{eq:Ehtrans}
E_{h}(t)&=&\frac{1}{2}(1-\sigma_{ds})\alpha e^{-\alpha t}\tilde{\ba}(0),\\
\label{eq:Eetrans} E_{e}(t)&=&-\frac{1}{2}(1+\sigma_{ds})\alpha
e^{-\alpha t}\tilde{\ba}(0).
\end{eqnarray}
In particular, for symmetric electron-hole bilayers, $\sigma_{ds}=0$,
these fields are equal and opposite. These predictions, along with the check
that the long-time electric fields in the two layers are indeed
independent of currents in individual layers (as long as the
current in the symmetric channel is constant), will provide robust
check of the on-set of excitonic condensation in these systems.

%-----------------------------------------------------------------------------%

\section{In-plane Josephson effect}
\label{sec:jj}

\begin{figure}[thbp]
\begin{center}
\includegraphics[width=2in]{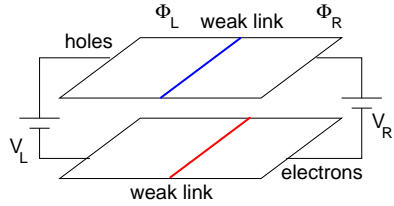}
\caption{(Color Online) In-plane Josephson effect in electron-hole bilayers.
Dipolar phases on two sides of the weak link are uniform, and the
time-evolution of the {\it dipolar phase difference} $\Phi_d$ is
determined by the voltage difference $V_L-V_R$.}
\label{fig:ehweaklink}
\end{center}
\end{figure}

Now we discuss transport across a weak link in the presence of voltage
applied across the link. This effect was proposed in the case of bilayer
quantum Hall systems a long time ago,~\cite{wen,ezawa,wen2} but it has not
yet been experimentally observed. Here we focus on bilayer electron-hole
systems with conventional tunneling across the weak links within each layer.
In a bilayer system with vanishingly small bare recombination rate, the dipolar
phase is fixed spontaneously across the sample and it varies
across the weak link with little energy cost. The action for the
dipolar phase $\Phi(\br,t)$ is given by
\begin{equation}
\label{eq:act} {\cal
S}=\int_{\br,t}\left[\frac{C}{2}\left(\partial_t\Phi-ea_0\right)^2-
\frac{\rho_d}{2}\left(\nabla\Phi-e\ba\right)^2\right]
\end{equation}
where $C$ is the capacitance, and
$a_0(\br,t)=a_e(\br,t)-a_h(\br,t)$ is the difference between
potentials $V_\alpha$ ($\alpha=L,R$) in the two layers, which
naturally couples to the time-derivative of the dipolar phase. We
consider a system where the dipolar phase is uniform on the two
sides and has spatial gradients only near the weak link
(Fig.~\ref{fig:ehweaklink}). In the regime $C\gg \rho_d$, the time
evolution of dipolar phase is given by
$\Phi_\alpha(t)=\Phi_{0\alpha}+e\int dt'V_\alpha(t')$. Therefore,
the dipolar phase {\it difference} evolves as
\begin{equation}
\label{eq:diff} \Phi_d(t)=\Phi_L(t)-\Phi_R(t)=\Phi_{0d}+e\int
dt'\left[V_L(t')-V_R(t')\right].
\end{equation}
In analogy with a Josephson junction, the dipolar current across
the weak link is given by $J(t)=J_c\sin\Phi_d(t)$. Since it is
possible to have a nonzero phase difference $\Phi_{0d}\neq 0$ in
the absence of interlayer voltage, the system can exhibit in-plane
(condensate) current across the weak link.

Now we can discuss various Josephson effects based on the
time-dependence of the applied voltage. We start with
$V_L-V_R=V_0+V_1\cos\omega t$, which leads to the phase-difference
evolution $\Phi_d(t)=\Phi_{0d}+eV_0t+(eV_1/\omega) \sin\omega t$.
Thus, for $V_1=0$, we get the ac Josephson effect, where the
in-plane current oscillates in the presence of a dc voltage,
$J(t)=J_c\sin(\Phi_{0d}+eV_0t)$. On the other hand, for an ac
voltage, we find that the Josephson current shows Shapiro steps at
frequencies $\omega_n=eV_0+n\omega$,
\begin{equation}
\label{eq:shapiro}
J(t)=J_c\sum_{n=-\infty}^{\infty}J_n(eV_1/\omega)\sin(\Phi_{0d}+\omega_nt)
\end{equation}
where weight of the current with frequency $\omega_n$ is given by
$J_n(eV_1/\omega)$, the Bessel function of order $n$. Let us assume that the
frequency $\omega$ is fixed, vary the amplitude $V_1$ of the applied voltage,
and focus on the first two harmonics, at frequencies
$\omega_0=\Phi_{0d}+eV_0$ and $\omega_1=\omega_0+\omega$. The strengths of
these two components are given by $J_0(eV_1/\omega)$ and $J_1(eV_1/\omega)$
respectively. Thus, as we sweep the voltage amplitude $V_1$, the frequencies
of the two harmonics are unchanged, but their strengths vary. In particular,
when $eV_1/\omega=0$, only the first harmonic contributes, whereas for
$eV_1/\omega\approx 2$ (the first zero of $J_0$), only the second harmonic
contributes and the ratio of these harmonics is 1.0:0.6. One can produce
the same Shapiro steps by applying an
in-plane ac magnetic field. The ac magnetic field produces an ac in-plane
electric field and would have the same effect as an ac voltage, as discussed
above.

We conclude this section with a brief derivation of the critical
current $J_c$ in terms of the microscopic Hamiltonian for
tunneling across the weak link. The tunneling Hamiltonian is given
by
\begin{equation}
\label{eq:jjtunnel} H_{t}=\int_{\br\in L,\br'\in
R}\left[T^{e}_{\br\br'}c^{\dagger}_{e\br} c_{e\br'} +
T^{h}_{\br\br'}c^{\dagger}_{h\br}c_{h\br'}+\mathrm{h.c.}\right].
\end{equation}
Here, $c^{\dagger}_{e\br}$ ($c^{\dagger}_{h\br}$) creates an
electron (hole) in the bottom (top) layer at position $\br$, and
$T^{e}$ ($T^{h}$) are the electron (hole) tunneling matrix
elements. The tunneling rate across the weak-link is given by
Fermi's golden rule, $\gamma\propto H_t^2$. However, in the limit
of vanishing potential difference across the link, the only term
which remains nonzero is
\begin{equation}
\gamma_0\propto T^{e}T^{h}\langle
c^\dagger_{eL}c_{eR}c^{\dagger}_{hL}c_{hR} \rangle \sim
T^{e}T^{h}\langle c^\dagger_{eL}c^{\dagger}_{hL}\rangle\langle
c_{eR}c_{hR}\rangle\sim T^{e}T^{h}\Delta_L\Delta^{*}_R.
\end{equation}
Thus the critical current $J_s\propto\gamma_0$ is proportional to {\it both}
electron and hole tunneling matrix elements and it is necessary to have weak
links in {\it both layers} for the in-plane Josephson effect to
occur.~\cite{cav} We emphasize that in conventional Josephson junctions in
superconductors, the tunneling matrix elements are typically spin independent 
and a systematic study of tunnel junctions in which the tunneling for the up
and down spins varies significantly has not been performed. On the other 
hand, bilayer electron-hole 
systems offer junctions where the tunneling for the two constituents of the
exciton (electrons and holes) can be independently controlled.

%-----------------------------------------------------------------------------%

\section{Noise in the in-plane current}
\label{sec:noise}

In the last two sections, we considered transport in a bilayer system when
the system is driven by external fields. In this section, we focus on
{\it noise} in the system when it is not driven. We consider the noise in
the in-plane current and show that it, combined with the measurement of
counterflow superfluidity, provides a direct measure of the superfluid
velocity of the excitonic condensate. Noise in any observable is a probe of
the excitation spectrum of the system. Therefore, in condensate state, we
expect the noise to probe the low-lying excitations, namely the collective
sound mode of the dipolar superfluid. To this end, let us consider noise
correlations between currents in the two layers. We start with the
symmetrized current-current correlator,
\begin{equation}
\label{eq:jj} C^{eh}_{ij}(\br,t)=\langle
\bJ_{ie}(\br,t)\bJ_{jh}(0)\rangle+\langle\bJ_{jh}(0)
\bJ_{ie}(\br,t)\rangle=2\Re\langle
T\bJ_{ie}(\br,t)\bJ_{jh}(0)\rangle
\end{equation}
where $i,j$ denote 2-dimensional Cartesian components of the
current density and  $T$ stands for time ordering. This correlator
corresponds to autocorrelation between current fluctuations at two probes
separated by $\br$ at times $t$ apart, when the system is in equilibrium
without any external voltage applied.
At low temperature, the quasiparticle contribution to the
fluctuations is vanishingly small, assuming a fully gapped
spectrum. Therefore, we can approximate the current fluctuations
as $\delta\bJ_{h(e)}=\pm e\rho_d\delta\nabla\Phi$, where the
dynamics of the dipolar phase $\Phi$ is governed by action
(\ref{eq:act}). It is straightforward to evaluate the time-ordered
current-current correlator in momentum space
\begin{equation}
\label{eq:jjmomentum} \langle
T\bJ^{*}_{i+}(\bk,\omega)\bJ_{j-}(\bk\omega)\rangle=(e\rho_d)^2k_ik_j
\frac{i}{C(\omega^2-v_c^2k^2)},
\end{equation}
where $v_c=\sqrt{\rho_d/C}$ is the velocity of the superfluid
sound mode. Therefore, the real-space current-current correlator
is given by
\begin{equation}
\label{eq:cij}
C^{eh}_{ij}(\br,t)=-\frac{(e\rho_d)^2}{Cv_c}\int^{\Lambda}\frac{d\bk}{(2\pi)^2}
\frac{k_ik_j}{k}\cos(\bk\cdot\br-v_ck|t|),
\end{equation}
where $\Lambda\sim r_s^{-1}$ is the momentum cutoff beyond which the
hydrodynamic description of the excitonic condensate fails. It
follows from Eq.(\ref{eq:cij}) that the current-current correlator
vanishes for $i\neq j$.

\begin{figure}[thbp]
\begin{center}
\includegraphics[width=2in]{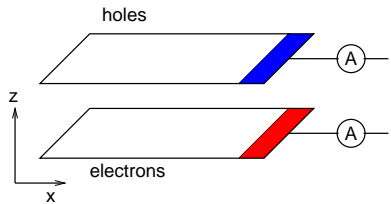}
\caption{(Color Online) 
Noise measurement in non-driven bilayer system. Current
fluctuations in the hole and electron layers are (anti)correlated
in the condensate state, where fluctuations are primarily due to
the superfluid sound mode. Therefore, the noise power spectrum
probes the properties of this sound mode.} \label{fig:ehnoise}
\end{center}
\end{figure}

In experiments, it is natural to consider the correlator for
(integrated) current fluctuations at the edge of the sample.
Consider, for example,
\begin{equation}
C^{eh}_{xx}(x,t)=\int dy
C^{eh}_{xx}(\br,t)=-\frac{(e\rho_d)^2}{Cv_c}\int dy\int
\frac{d\bk}{(2\pi)^2}k_x^2\cos(\bk\cdot\br-v_ck|t|).
\end{equation}
The integration along the $y$-axis only retains the $k_y=0$
component and the noise correlator becomes
\begin{equation}
\label{eq:cxx}
C^{eh}_{xx}(x,t)=-\frac{(e\rho_d)^2}{Cv_c}\int^{\Lambda}_{-\Lambda}
\frac{dk_x}{(2\pi)}|k_x|\cos(k_xx-v_c|k_xt|).
\end{equation}
It is straightforward to evaluate the integral and we get
\begin{equation}
\label{eq:cxxfinal}
C^{eh}_{xx}(x,t)=-\frac{(e\rho_d\Lambda)^2}{Cv_c}\left[\frac{\sin(x_{-}
\Lambda)}{x_{-}\Lambda}+\frac{\sin(x_{+}\Lambda)}{x_{+}\Lambda}-
\frac{(1-\cos(x_{-}\Lambda))}{(x_{-}\Lambda)^2}
-\frac{(1-\cos(x_{+}\Lambda))}{(x_{+}\Lambda)^2}\right],
\end{equation}
where we have defined auxiliary variables $x_{\pm}=(x\pm v_ct)$.

Eq.(\ref{eq:cxxfinal}) has several interesting features. The
correlator is only a function of $x_\pm=(x\pm v_ct)$ and is
symmetric in them. This is expected since the dipolar action,
Eq.(\ref{eq:act}), is Lorentz invariant in the absence of external
fields. When the two probes measure total current
fluctuations at the same co-ordinate, $x=0$ and $x_\pm=\pm v_c t$,
the correlator simplifies to
\begin{equation}
C^{eh}_{xx}(0,t)=-2e^2\Lambda^2\rho_d v_c\left[\frac{\sin(v_c\Lambda
t)} {v_c\Lambda t}-\frac{(1-\cos(v_c\Lambda t))}{(v_c\Lambda
t)^2}\right].
\end{equation}
This correlator has a power spectrum
\begin{equation}
\label{eq:cxxpower} C^{eh}_{xx}(\omega)=-\frac{2\pi
e^2\rho_d}{v_c}|\omega|\theta(v_c\Lambda- |\omega|).
\end{equation}

Thus, the power spectrum of the current noise is linear with a slope which
is proportional to the superfluid density and inversely proportional to the
superfluid collective-mode velocity $v_c$. Eq.(\ref{eq:cxxpower}), combined
with the counterflow superfluidity result, $J_d=2e^2\rho_ddB_{||}$, provides
a {\it direct measurement} of the superfluid velocity $v_c$. The power
spectrum vanishes
beyond $\omega_0=v_c\Lambda\propto v_cr_s^{-1}$ because the hydrodynamic
description is not valid beyond momentum $\Lambda$ and therefore it cannot
probe frequencies higher than $v_c\Lambda$.
A similar calculation of current fluctuations within a single layer gives
$C^{ee}_{ij}(\br,t)=C^{hh}_{ij}(\br,t)=-C^{eh}_{ij}(\br,t)$. This
result is crucially based on the system being in the condensate
state. We emphasize that this result will not hold in bilayer quantum Hall
systems in the phase-coherent state. When the two layers are weakly coupled,
current fluctuations in the two layers will have qualitatively different
nature. Thus, measurement of the in-plane current noise can
provide yet another signature of the excitonic condensate state
and it's power spectrum can provide a measure of the superfluid
collective mode velocity and it's evolution near the phase
boundary.

%-----------------------------------------------------------------------------%

\section{Discussion}
\label{sec:disc}
We have considered the transport and noise in dipolar excitonic condensate
in electron-hole bilayers. These bilayers are a promising candidate for
the realization of excitonic condensate. In principle, the condensation of
excitons is not well-defined because excitons are metastable bound states of
electron-hole pairs which eventually decay. However, with present
semiconductor heterostructures, it possible to create electron-hole bilayers
with very low recombination rates and therefore probe properties of the
excitonic condensate without destroying it. We find that the dipolar phase
has a number of nontrivial transport properties that can lead to the
condensate detection.

We have discussed the drag and counterflow transport features of these
bilayers taking into account the effect of condensate and quasiparticles.
We find that for the counterflow setup, the steady-state current is carried
solely the condensate and hence there is no voltage drop in either layer, in
the ideal case. In contrast, for the drag experiment, we find that the
electric fields in both layers are the same. These two results are specific
cases of our primary result, Eqs.(\ref{eq:Eh}) and (\ref{eq:Ee}), which
shows that the electric fields in the two layers are the same irrespective
of the current distributions in the electron and the hole layers, and is
determined only by their sum total.~\cite{vig} We also discussed the analog of
in-plane Josephson effect and showed that the existence of weak links in
{\it both} layers is instrumental to it.

We have demonstrated how noise spectroscopy of in-plane current fluctuations
in the electron-hole bilayers can provide a signature of the condensate state
as well as a direct measurement of the collective mode velocity $v_c$.
The current fluctuations at two probes, one in the electron layer and
other in the hole layer, will be naturally correlated if ground state
consists of excitons and the low-lying excitations are superfluid collective
modes. On the other hand, if the two layers are uncoupled, these fluctuations
will be uncorrelated. We find that the power spectrum of the current noise
is linear with slope $-2\pi e^2\rho_d/v_c$. The noise spectrum, combined with
the measurement of dipolar supercurrent $J_d=2e^2\rho_d dB_{||}$, will allow
us to extract information about both the dipolar phase stiffness $\rho_d$
and the collective mode velocity $v_c$. This
power spectrum will change significantly when the bilayer system undergoes
a quantum phase transition with increasing $d/r_s$, going from excitonic
condensate
to weakly coupled layers. We emphasize that the noise spectroscopy is
complementary to the transport experiments. In transport experiments, the
bilayer system is perturbed using external fields and the response of
excitonic condensate is measured. The noise spectroscopy, on the other hand,
is performed on a bilayer system in equilibrium, where statistical and
quantum fluctuations provide the probe of low-lying excitations of the
system. These complimentary measurements will provide various signatures
of the superfluid condensate and deepen our understanding of excitonic
condensates in semiconductors.

It is a pleasure to acknowledge discussions with Peter Littlewood. The work
at LANL was supported by the DOE. Sandia National Laboratories is a
multi-program laboratory operated by Sandia Corporation, a Lockheed-Martin
Company, for the U. S. Department of Energy under Contract No.
DE-AC04-94AL85000.

%-----------------------------------------------------------------------------%
% notes on dimensions used in the paper:
% I have h=1=c. Thus [c]=[c^]=1/length. [\Delta]=1/Area.
% I get [\varphi]=1/q, [\rho_d]=Energy, [A]=1/length.q
% [J]=q/time.length (note the extra \delta-function in df/dA.
% [1/C]=Energy.Area, [v]=[\sqrt{\rho/C}]=Energy.length=length/time.
%-----------------------------------------------------------------------------%

%-----------------------------------------------------------------------------%

\begin{thebibliography}{99}
\bibitem{keldysh} L.V. Keldysh, in {\it Bose-Einstein Condensation}, edited
by A. Griffin {\it et al.} (Cambridge University Press, Cambridge,
1995) and reference therein.
\bibitem{bilayer} Yu.E. Lozovik and V.I. Yudson,
Pis'ma Zh. Eksp. Teor. Fiz.{\bf 22}, 556 (1975) [JETP Lett. {\bf
22}, 274 (1975)]; Solid State Commun. {\bf 19}, 391 (1976); Zh.
Eksp. Teor. Fiz. {\bf 71}, 738 (1976) [Sov. Phys. JETP {\bf 44},
389 (1976)].
\bibitem{rice} T.M. Rice, Solid State Phys. {\bf 32}, 1 (1977);
J.C. Hensel {\it et al.}, {\it ibid.} {\bf 32}, 88 (1977).
\bibitem{shev} S.I. Shevchenko, Phys. Rev. Lett. {\bf 72}, 3242 (1992).
\bibitem{vig} G. Vignale and A.H. MacDonald, Phys. Rev. Lett. {\bf 76}, 2786
(1996).
\bibitem{palo} S.De Palo, F. Rapisarda, and Gaetano Senatore, 
Phys. Rev. Lett. {\bf 88}, 206401(2002).
\bibitem{sivan} U. Sivan, P.M. Solomon, and H. Shtrikman, Phys. Rev. Lett. 
{\bf 68}, 1196 (1992).
\bibitem{kane} B.E. Kane, J.P. Eisenstein, W. Wegscheider, L.N. Pfeiffer, 
and K.W. West, App. Phys. Lett. {\bf 65}, 3266 (1994).
\bibitem{vij} S. Vijendran, P.J.A. Sazio, H.E. Beere, G.A.C. Jones, D.A. 
Ritchie, and C.E. Norman, J. Vac. Sci. Technol. B {\bf 17}, 3226 (1999).
\bibitem{pohlt}M. Pohlt, M. Lynass, J.G.S. Lok, W. Dietsche, K.v. Klitzing, 
K. Eberl, and R. M\"{u}hle, App. Phys. Lett. {\bf 80}, 2105 (2002).
\bibitem{ayp}A.V. Balatsky, Y.N. Joglekar, and P.B. Littlewood, Phys. Rev.
Lett. {\bf 93}, 266801 (2004).
\bibitem{3he} D. Vollhardt and P. W\"{o}lfle, {\it The Superfluid Phases
of Helium 3} (Taylor and Francis, London, 1990).
\bibitem{jpe} J.P. Eisenstein and A.H. MacDonald, Nature (London) {\bf 432},
691 (2004) and references therein.
\bibitem{shay} E. Tutuc, M. Shayegan, and D. Huse, Phys. Rev. Lett. {\bf 93},
036802 (2004).
\bibitem{butov} L.V. Butov, A.C. Gossard, and D.S. Chemla, Nature (London)
{\bf 418}, 751 (2002).
\bibitem{snoke} D. Snoke, S. Denev, Y. Liu, L.N. Pfeiffer, and K.W. West,
Nature (London) {\bf 418}, 754 (2002); D. Snoke, Science {\bf 298}, 1368 
(2002).
\bibitem{lai} C.W. Lai, J. Zoch, A.C. Gossard, and D.S. Chemla, Science
{\bf 303}, 503 (2004).
\bibitem{voros} Z. V\"{o}r\"{o}s, R. Balili, D.W. Snoke, L.N. Pfeiffer, and 
K.W. West, Phys. Rev. Lett. {\bf 94}, 226401 (2005).
\bibitem{rojo} See, for example, A.G. Rojo, J. Phys. {\bf 11}, R31 (1999) 
and references therein. 
\bibitem{wen} X.G. Wen and A. Zee, Phys. Rev. Lett. {\bf 69}, 1811 (1992).
\bibitem{ezawa} Z.F. Ezawa and A. Iwazaki, Int. J. Mod. Phys. B {\bf 6}, 3205
(1992).
\bibitem{wen2} X.G. Wen and A. Zee, Europhys. Lett.{\bf 35}, 227 (1996).
\bibitem{cav} To our knowledge, the necessity of having weak links in both
layers has not been addressed explicitly in the literature.
\end{thebibliography}
\end{document}